\documentstyle[twoside,11pt]{article}
\def\greaterthansquiggle{\raise.3ex\hbox{$>$\kern-.75em\lower1ex\hbox{$\sim$}}}
\def\lessthansquiggle{\raise.3ex\hbox{$<$\kern-.75em\lower1ex\hbox{$\sim$}}}
\newcommand{\beq}{\begin{equation}}
\newcommand{\eeq}{\end{equation}}
\newcommand{\beqa}{\begin{eqnarray}}
\newcommand{\eeqa}{\end{eqnarray}}
\newcommand{\beqan}{\begin{eqnarray*}}
\newcommand{\eeqan}{\end{eqnarray*}}
\newcommand{\ba}{\begin{array}}
\newcommand{\ea}{\end{array}}

\newcommand{\ra}{\rightarrow}

\newcommand{\ve}{\varepsilon}
\newcommand{\vp}{\varphi}

\newcommand{\wt}{\widetilde}

\newcommand{\F}{{\cal F}}

\newcommand{\R}{{\cal R}}

\def\nz{\ifmmode {I\hskip -3pt N} \else {\hbox {$I\hskip -3pt N$}}\fi}
\def\zz{\ifmmode {Z\hskip -4.8pt Z} \else
       {\hbox {$Z\hskip -4.8pt Z$}}\fi}
\def\qz{\ifmmode {Q\hskip -5.0pt\vrule height6.0pt depth 0pt
       \hskip 6pt} \else {\hbox
       {$Q\hskip -5.0pt\vrule height6.0pt depth 0pt\hskip 6pt$}}\fi}
\def\rz{\ifmmode {I\hskip -3pt R} \else {\hbox {$I\hskip -3pt R$}}\fi}
\def\cz{\ifmmode {C\hskip -4.8pt\vrule height5.8pt\hskip 6.3pt} \else
       {\hbox {$C\hskip -4.8pt\vrule height5.8pt\hskip 6.3pt$}}\fi}

\def\au{{\setbox0=\hbox{\lower1.36775ex%
\hbox{''}\kern-.05em}\dp0=.36775ex\hskip0pt\box0}}
\def\ao{{}\kern-.10em\hbox{``}}

\normalsize
\begin{document}
\bibliographystyle{plain}
\begin{titlepage}
\vspace*{2cm}
\begin{center}
{\Large \bf Trapped Surfaces in Vacuum Spacetimes}\\[40pt]
R. Beig* \\
Institut f\"ur Theoretische Physik \\
Universit\"at Wien \\
e-mail BEIG@AWIRAP.BITNET \\[30pt]
N. \'O Murchadha* \\
Physics Department, University College \\
Cork, Ireland \\
e-mail STEP8030@IRUCCVAX.UCC.IE
\vfill
{\bf Abstract}
\end{center}

An earlier construction by the authors of sequences of globally regular,
asymptotically flat initial data for the Einstein vacuum equations
containing trapped surfaces for large values of the parameter is
extended, from the time symmetric case considered previously, to the case
of maximal slices. The resulting theorem shows rigorously that there
exists a large class of initial configurations for non-time symmetric
pure gravitational waves satisfying the assumptions of the Penrose
singularity theorem and so must have a singularity to the future.
\vfill
\noindent *) Supported by Fonds zur F\"orderung der wissenschaftlichen
Forschung in \"Osterreich, Projekt Nr. P9376-PHY.
\end{titlepage}

\section{Introduction}

This paper is concerned with the existence of maximal, asymptotically
flat initial data for the Einstein vacuum equations which contain trapped
surfaces. The significance of this problem stems from the singularity
theorems of General Relativity -- or rather, the issues they raise.
Take the Penrose theorem [1] for example. This shows that a spacetime
$N$ with reasonable matter (in particular: a vacuum spacetime) evolving
from a non-compact (in particular: asymptotically flat) Cauchy surface
$\wt M \subset N$ cannot be null-geodesically complete when $\wt M$
contains a trapped surface (TS). The theorem still holds when `trapped
surface' is replaced by `outer trapped surface'. By a TS we shall
always mean an outer trapped surface. An obvious question to ask, is under
what circumstances the TS-condition is actually satisfied. From the
point of view of the initial-value problem this question really consists
of two parts:
\begin{enumerate}
\item Can one find conditions on $\wt M$, e.g. in terms of `free data',
in order for $\wt M$ to contain a TS?
\item What are conditions on $\wt M$ so that there is a TS in a
Cauchy development of $\wt M$?
\end{enumerate}
Thirdly, since the singular behaviour predicted by the singularity theorems
is merely the existence of incomplete causal geodesics, one would like
to understand the local and global nature of these `singularities'. In
particular, one would like to know
\begin{enumerate}
\item[3.] Does $N$ satisfy some version of the cosmic censorship
hypothesis (CCH)?
\end{enumerate}

Clearly, question 3.) is the most interesting one. Since 2.) and 3.)
are by their nature questions concerning the global Cauchy problem (see
Chrusciel [2] for a review of what is known in this area) they are also
very difficult. Only in the case of gravity coupled to a massless
scalar field, and with spherical symmetry imposed on the geometry and
the source, has a full analysis of these issues  been achieved in the
work of Christodoulou [3].

The present paper is devoted to the study, for initial data
corresponding to pure gravitational radiation, question 1.) in the
above list. Although this is a more modest task than attacking 2.)
or 3.), it is widely assumed that a full understanding of 1.) should
also shed light on 3.). The reason for this lies in the belief, which
is of course just one aspect of CCH, that the maximal Cauchy
development of an asymptotically flat slice $\wt M$ with trapped
surfaces is a black hole spacetime. In particular, singularities should
be inside the apparent horizon which, in turn, is surrounded by an event
horizon. No counterexamples are known to this `event horizon conjecture'
(see Israel [4] for steps twards a resolution). Initial data like the
ones constructed in this paper are thus considered to be relatively
`safe' from the viewpoint of CCH. At the other extreme there have been
studies of families of initial data the limiting members of which are
singular, but without TS's. Such families, found numerically by
Nakamura et al. [4], mimick the formation of naked singularities
 which Shapiro and Teukolsky [5] obtain in their
simulations of collapsing localized dust. Since these solutions have
only axial symmetry, the singularities are thought to be a much more
serious threat to CCH than previously known counterexamples all of
which are spherical. Sequences analogous to the ones in [4], but for
the vacuum case, which are in a sense complementary to the vacuum
sequences considered in our previous work [6], have recently been
found numerically by Abrahams et al. [7] (see also O'Murchadha [8]).
These may be taken as possible evidence for the violation of CCH even
in the vacuum case, but the issue is completely open.

There has been considerable interest recently in conditions on Cauchy
data in order to have trapped surfaces.
Only the spherically symmetric case is fully understood so far [9].
For generalizations of these
results, mostly motivated by Thorne's Hoop Conjecture, see Malec [10]
and Flannagan [11]. For a completely different approach, based on a
suggestion by Penrose, see Tod [12] and Barrabes et al. [13].
(Compare, also, the results of Schoen and Yau [14], together with
O'Murchadha [15].)

In our work [6] we have shown the existence of time symmetric,
asymptotically flat, vacuum initial data with TS's, thus settling a
question raised by Bartnik [16]. (For earlier numerical evidence see
Eppley [17] and Miyama [18].)  Our construction is based in an
essential manner on the conformal approach (see Choquet-Bruhat and
York [19]) to solving the  constraints of General
Relativity. In the time-symmetric case
this consists of picking a 3-metric $\wt g$ on $\wt M$ which tends to a
fixed flat metric $\delta$ near infinity and seeking as the physical
metric a metric $g'$ in the conformal class defined by $\wt g$, also
tending to $\delta$ at infinity. This gives rise to an elliptic equation
for the conformal factor $\wt \Phi$, the Lichnerowicz equation, together with
the boundary condition that $\wt \Phi$ tends to 1 at infinity. Some
condition on $\wt g$ has to be satisfied in order for there to be a
solution of this problem. This condition has the form of an inequality
on a global conformal invariant associated with $\wt g$, the Yamabe
number $Y$ (Lee and Parker [20], see also O'Murchadha [21]), namely
$Y > 0$. We now choose a sequence of metrics $\wt g_n$, called a critical
sequence (CS), for which $Y > 0$, but tending as $n \ra \infty$ to a
metric for which $Y = 0$. It follows that the ADM masses $m_n$
increase without bound as $n \ra \infty$. It furthermore can be shown
that higher-order multipoles diverge no more rapidly than the mass
for large $n$.
{}From this, in turn, one can infer that TS's form. The present work
extends the above construction to the maximal (non-time symmetric) case,
provided the suitably rescaled extrinsic curvature stays bounded as $n$
increases. Then the ADM mass is dominated by its time symmetric part,
so that essentially the same conclusion obtains as before.
In the context of the discussion at the beginning of this section some
features of the above scenario are worth pointing out. Firstly,
for large $n$, CS's can {\bf not} result from time evolution via the
Einstein equations. To see this, recall that maximal, asymptotically
flat slices in a single spacetime asymptotically have to be boosts
or translations of each other (see Bartnik et al. [22]). We now
invoke the fact that the spatial metrics $\wt g$ in our construction
are, in addition to being asymptotically flat at infinity to leading
order, asymptotically spherical (``Schwarzschildian'') to the next
order. This implies that
asymptotic boosts are no longer allowed, and we are left with the
possibility of slices which are mapped into each other by asymptotic
translations. But then the mass $m$ has to be conserved which
contradicts $\lim_{n \ra \infty} m_n = \infty$.
This suggests that one way to improve our results would be to try to
establish a universal bound on $Y$, below which $g'$ has TS's so that
it would no longer be necessary to consider sequences with $Y \ra 0$.
Secondly, it would be interesting to have a better understanding of
$Y$ in terms of conformal geometry on the local or quasilocal level
and use this to study possible analogues of the Hoop Conjecture in
the vacuum case.

The critical sequences discussed in [6] are explicitly time-symmetric.
This means that the TS's which form are both future and past trapped.
This, in turn, means that the spacetimes evolving from these data for
large $n$ are both future- and past-incomplete. Thus, if cosmic
censorship holds, we should have a configuration which starts out
as a white hole in the distant past, expands to reach an instant of
time-symmetry and then collapses to form a black hole in the distant
future. In contrast, in our current work we break the time-symmetry
assumption on the initial hypersurface by introducing some nonzero
(albeit relatively small) extrinsic curvature. Thus, as we move along
a critical sequence, we do not expect future and past apparent horizons
to appear simultaneously. Of course, if we move sufficiently far along
any critical sequence the effects of the ever-increasing ADM mass will
swamp the extrinsic curvature and we return to the time-symmetric
situation with past as well as future singularities. Still, along any
critical sequence we should have a band of data, solutions to the
constraints, with a future apparent horizon but no past apparent
horizon (if they appear in the opposite order, just multiply the
extrinsic curvature by $-1$). These data now form possible candidates
to generate spacetimes which represent the inflow of gravitational
radiation from past null infinity of such an amount as to cause
gravitational collapse and the formation of a black hole. We realize
that the absence of a past horizon in an initial data set is no
guarantee of regularity into the past (see question 2.) of our first
paragraph). However, it certainly is a necessary condition. Thus we can
only claim that it is plausible that at least some of these initial
data sets that are singular to the future, are regular to the past.

Recall that in the conformal approach to solving the constraints [19]
in the maximal case we have, in addition to the `background' metric
$\wt g_{ab}$, a background extrinsic curvature $\wt p_{ab}$ related to
$p'_{ab}$, the physical extrinsic curvature, by $p'_{ab} = \wt \Phi^{-2}
\wt p_{ab}$.
In the calculation presented here we move the background geometry
along a critical sequence while
keeping the background extrinsic curvature bounded. The obvious
alternative approach is to hold $\wt g$ fixed and
scale up the extrinsic curvature $\wt p$. While there are obvious technical
difficulties we expect that this approach would also work and give
initial data which are significantly more time-asymmetric than those
produced here. This approach deserves further study.

We restrict our attention to maximal initial data. The maximal condition
allows a separation of the momentum constraint from the Hamiltonian
constraint and reduces the problem of constructing initial data from a
quasi-linear coupled system to the problem of solving a single
quasi-linear elliptic equation. We probably can handle the situation
where the trace of the extrinsic curvature, while not vanishing, is small.
Unfortunately, a systematic technique along the lines of this paper
giving TS's in the non-maximal case requires
a better understanding of the coupling between the Hamiltonian and
momentum constraint than is presently available.

Our paper is organized as follows. In Sect. 2 we recall the conformal
approach to solving the Hamiltonian constraint in the maximal case
and explain our basic assumptions. As in [6] we use a technique
based on conformal compactification of 3-space $\wt M$. On the
resulting compact manifold $M$ with metric $g$ we use for simplicity,
instead of the conformal invariant $Y$ the quantity $\lambda_1$, the
lowest eigenvalue of the conformal Laplacian on $(M,g)$. The latter
is conformally non-invariant, but is positive (zero) iff $Y$ is
positive (zero), and this sign condition is the only one we use.
In Sect. 3 the notion of critical sequence is introduced and the
theorem on TS-formation is proved. In the Appendix we prove an
existence and uniqueness theorem for the Lichnerowicz equation
appropriate for our setting.

\section{Solving the Hamiltonian Constraint}

The basic formula underlying the standard [19] method of solving
the Hamiltonian constraint is as follows. Define the conformal
Laplacian associated with the 3-metric $g_{ab}$ by
$$
L_g := - \Delta_g + \frac{1}{8} {\cal R} [g],   \eqno(2.1)
$$
$\Delta_g := g^{ab} D_a D_b$ being the standard Laplacian acting
on scalars and ${\cal R} [g]$ the scalar curvature of $g$. Then,
if we put
$$
g'_{ab} = \omega^4  g_{ab}, \qquad
\Phi' = \omega^{-1} \Phi \qquad \omega > 0, \eqno(2.2)
$$
we find that
$$
L_{g'} \Phi' = \omega^{-5} L_g \Phi. \eqno(2.3)
$$
If in particular $\Phi > 0$ and we set $\omega = \Phi$, we have
$\Phi' \equiv 1$ and there results the identity
$$
\frac{1}{8} \R[g'] = \Phi^{-5} L_g \Phi. \eqno(2.4)
$$
Thus $\R[g']$ is zero if and only if $\Phi$ satisfies the `linear
Lichnerowicz equation' $L_g \Phi = 0$. Recall that $\R = 0$ is just
the initial-value constraint in the time-symmetric case.

More generally, let us be given a trace-free, divergence-free w.r.t. to
$g_{ab}$ (TT-) tensor $p_{ab}$, i.e.
$$
g^{ab} p_{ab} = 0, \qquad D^a p_{ab} = 0. \eqno(2.5)
$$
Notice that Equ.-s (2.5) are conformally covariant if $p_{ab}$ is
transformed like
$$
p'_{ab} = \omega^{-2} p_{ab}. \eqno(2.6)
$$
Suppose now that a positive scalar $\Phi$ is subject to the equation
$$
L_g \Phi = \frac{1}{8} \Phi^{-7} p_{ab} p^{ab}, \qquad \Phi > 0,
\eqno(2.7)
$$
then the pair ($g'_{ab} = \Phi^4 g_{ab}$, $p'_{ab} = \Phi^{-2}
p_{ab}$) satisfies
$$
\R [g'] = \Phi^{-5} \Phi^{-7} p_{ab} p^{ab} =
g'{}^{ac} g'{}^{bd} p'_{ac} p'_{bd} \eqno(2.8)
$$
and
$$
D'{}^a p'_{ab} = 0
$$
that is to say the initial-value constraints of General Relativity in
the maximal, source-free case. The set of pairs $(g_{ab},p_{ab})$,
where $p_{ab}$ is TT, subject to the equivalence relation $g_{ab} \sim
g'_{ab}$ and $p_{ab} \sim p'_{ab}$ with $g'_{ab} = \omega^4
g_{ab}$ and $p'_{ab} = \omega^{-2} p_{ab}$ for some $\omega > 0$,
are called the `free data' for vacuum G.R. The word `free' here is
misleading for two reasons. Firstly, $p_{ab}$ being TT comes from solving
a system of partial differential equations. We will comment on how this
can be done. Secondly, $g_{ab}$ has to be such that the Lichnerowicz
equation (2.7) has a positive solution $\Phi$ with the required global
properties given the appropriate global properties of $(M,g_{ab},p_{ab})$.
To these we now turn.

We want the `physical' data $g'_{ab}$, $p'_{ab}$ to be in a suitable
sense asymptotically flat on some physical manifold $\wt M$. Our idea
here is to obtain $(\wt M,g',p')$ by `conformal decompactification'
from $(M,g,p)$ where $M$ is a compact manifold. Recall that it is often
convenient to
describe asymptotic flatness of a Riemannian manifold $(\wt M,g')$
by the existence of a manifold $M = \wt M \cup \{\Lambda\}$,
$\Lambda$ being a point, and a function $\Omega$ on $M$, going to zero at
$\Lambda$ in such a manner that  $g_{ab} = \Omega^2 g'_{ab}$
extends in a sufficiently regular way to a metric on $M$.
But this transition from
$\wt M$ to $M$ has a drawback: whereas in the physical picture we have
metrics, in the unphysical picture, due to the gauge freedom in
$\Omega$, we only have conformal metrics, and these are more complicated
to deal with. In the present situation, however, it is precisely the
conformal structure associated with $g_{ab}$ which is among the basic
objects, so the above freedom is in fact desirable. Thus we take $M$
to be compact and $g_{ab}$ a smooth metric thereon. Pick a point
$\Lambda \in M$, the point-at-infinity. (The results of the present
section could easily be generalized to the case where there is an
arbitrary but finite number of points-at-infinity.) We next define what
we mean by an asymptotic distance function (ADF) w.r. to $\Lambda$.
$\Omega^{1/2}$ is called an ADF, provided $\Omega$ is smooth on $M$,
positive outside $\Lambda$ and satisfies ($\Omega_a := D_a \Omega$,
$\Omega_{ab} := D_a D_b \Omega$, a.s.o.)
$$
\left. \Omega\right|_\Lambda = 0, \qquad
\left. \Omega_a\right|_\Lambda = 0, \qquad
\left. (\Omega_{ab} - 2g_{ab})\right|_\Lambda = 0,
\qquad \left. \Omega_{abc}\right|_\Lambda = 0. \eqno(2.9)
$$
For example, $\Omega$ could be taken to be the square of the geodesic
distance from $\Lambda$. Given $g_{ab}$ and $\Omega$, the metric
$\wt g_{ab}$, defined by $\wt g_{ab} = \Omega^{-2} g_{ab}$ is easily
seen to be asymptotically flat on $\wt M$ and have vanishing ADM mass.
More precisely, if $x^a$ are coordinates centered at $\Lambda$, in which
$\left. g_{ab}\right|_\Lambda = \delta_{ab}$ and $\left.\partial g_{ab}
\right|_\Lambda = 0$, then $\wt x^a = \Omega^{-1} x^a$ are coordinates in
a punctured neighbourhood of $\Lambda$, in which
$$
\wt g_{ab} = \delta_{ab} + O(\Omega), \eqno(2.10)
$$
$\delta_{ab}$ being theKronecker delta. Thus in order for $\Phi^4 g_{ab}$
to be an asymptotically flat metric the correct boundary
condition for $\Phi$ is something like
$$
\Phi = \Omega^{-1/2} + \frac{m}{2} + O(\Omega^{1/2}), \qquad
m = \mbox{constant}. \eqno(2.11')
$$
In fact, we shall have to weaken (2.11$'$) by only requiring that
$$
\Phi = \Omega^{-1/2} + \frac{m}{2} + O(\Omega^{\ve/2}),
\qquad 0 < \ve < 1, \eqno(2.11)
$$
$$
\partial \Phi = \partial(\Omega^{-1/2}) + O(\Omega^{(\ve-1)/2}).
\eqno(2.12)
$$
When $\Phi$ obeys (2.11,12), then $g'_{ab} = \Phi^4 g_{ab}$, in the
$\wt x^a$-coordinates, satisfies the asymptotic relations
$$
g'_{ab} = \left( 1 + \frac{m\Omega^{1/2}}{2}\right) \delta_{ab} +
O(\Omega^{(\ve+1)/2}), \eqno(2.13)
$$
$$
\partial g'_{ab} = \partial \left( 1 + \frac{m \Omega^{1/2}}{2}\right)
\delta_{ab} + O(\Omega^{\ve/2}). \eqno(2.14)
$$
Writing $\wt r^2 = \delta_{ab} \wt x^a \wt x^b$, we have
$$
\Omega = \wt r^{-2} + O(\wt r^{-4}) \quad \mbox{near } \Lambda
\eqno(2.15)
$$
so (2.13,14) are in fact the standard conditions for an asymptotically
Schwarzschildian 3-metric. The boundary conditions for $p'_{ab}$,
formulated in terms of $p_{ab}$, are as follows:
$$
p_{ab} = O(\Omega^{-9/4 + \ve/4}), \eqno(2.16)
$$
$$
\partial p_{ab} = O(\Omega^{-11/4 + \ve/4})
$$
which implies that $p'_{ab} = \Phi^{-2} p_{ab}$ satisfies
$$
p'_{ab} = O(\Omega^{3/4 + \ve/4}) = O(\wt r^{-3/2 - \ve/2}),
\eqno(2.17)
$$
$$
\partial  p'_{ab} = O(\wt r^{-5/2 - \ve}).
$$
The connection of the above `unphysical' formulation with the standard
physical formulation of the Lichnerowicz equation is as follows. We
define fields on $\wt M$
$$
\wt \Phi = \Omega^{1/2} \Phi, \qquad \wt p_{ab} = \Omega p_{ab},
\eqno(2.18)
$$
so that $\wt \Phi \ra 1$ at infinity and $\wt p_{ab}$ has the same
asymptotic behaviour as $p'_{ab}$ (Equ. (2.17)). Furthermore
$$
L_{\wt g} \wt \Phi = \frac{1}{8} \wt \Phi^{-7} \wt p_{ab} \wt p^{ab}
\qquad \mbox{on } \wt M. \eqno(2.19)
$$
The condition that the extrinsic curvature falls off faster than
$\wt r^{-3/2}$ is exactly the one that guarantees that the ADM mass be
finite and well-defined [24]. Naively, to leading order
the physical Lichnerowicz equation looks like
$\wt \Delta \wt \Phi + \frac{1}{8} \wt p_{ab} \wt p^{ab} = 0$ with
$\wt p_{ab} \wt p^{ab} = O(\wt r^{-3-\ve})$. Hence
$\wt \Phi = 1 + m/2 \wt r + O(\wt r^{-1-\ve})$. This gives us the standard
Schwarzschildian behaviour at infinity.

Using the results of Chaljub-Simon [23] one can construct many $p_{ab}$'s
satisfying (2.5) and the boundary conditions (2.16) in the case where
$M \cong S^3$ $(\wt M \cong {\bf R}^3)$.

We now come to the issue of existence of solutions $\Phi > 0$, satisfying
$$
L_g \Phi = \frac{1}{8} \Phi^{-7} p_{ab} p^{ab} \quad \mbox{on }
M \setminus \Lambda \eqno(2.20)
$$
together with (2.11,12).

First recall that $L_g$ is an essentially self-adjoint operator on
$C^\infty(M)$ w.r. to the standard $L^2$-inner product associated with
the metric $g$. In addition, the spectrum of $L_g$ is a pure point
spectrum and is bounded from below. Let $\lambda_1(g)$ be its lowest
eigenvalue.

\paragraph{Theorem:} Let $\lambda_1(g) > 0$. Then (2.20), (2.11,12)
have a unique solution which is smooth on $M \setminus \Lambda$.

\paragraph{Proof:} It is known (see e.g. [20]) that the problem can be
uniquely solved when $p_{ab}$ is zero. Call the corresponding solution
$G$ and write
$$
\Phi = G + h. \eqno(2.21)
$$
Then (2.20) takes the form
$$
L_g h = \frac{1}{8} (1 + G^{-1}h)^{-7} G^{-7} p_{ab} p^{ab}, \eqno(2.22)
$$
where $G^{-1}h$ simply means pointwise multiplication.
Note that the r.h. side of (2.22) diverges at the point $\Lambda$,
but does so in a relatively mild fashion, i.e. like
$O(r^{-2+\ve})$, $0 < \ve < 1$. This is enough for the existence theorem
proven in the Appendix to hold.

\section{Critical Sequences}
Similarly as in [6] we call a sequence $g_n$ of 3-metrics on $M$ with
$\lambda_1(g_n) > 0$ critical when it converges uniformly together with
all its derivatives to a metric $g_\infty$ with $\lambda_1(g_\infty) = 0$.
We also require a critical sequence (CS) to be such that the metrics
$g_n$ and the connections $\partial g_n$ all coincide at the point
$\Lambda$, so that an ADF $\Omega^{1/2}$ can be found which is independent
of $n$. We furthermore assume that for each $n$ we are given a TT-tensor
$p_{ab}$
which is smooth in $M \setminus \Lambda$ and satisfies (2.16), where
the constants involved in the $O$-symbols are understood to be
independent of $n$. The existence of such a sequence of $p_{ab}$'s,
in the case where $M \cong S^3$,
follows from the work of Chaljub-Simon [23].
(Basically the reason for this is that the criticality of the limiting
metric $g_\infty$ plays no role for the existence of a TT-tensor relative
to $g_\infty$.)
For typographical reason
we omit the index ``$n$'' in the following expressions. Since
$\lambda_1(g) > 0$, the operator $L_g$ has a Green `function' $G(x,x')$.
The function $G(x)$ solving the time-symmetric constraints is, of course,
$$
G(x) = G(\Lambda,x). \eqno(3.1)
$$
Thus the solution to Equ. (2.22) can be written as
$$
h(x) = \frac{1}{8} \int_M G(x,x') [1+(G^{-1}h)(x')]^{-7}
(G^{-7} p_{ab} p^{ab})(x') dV_g(x'). \eqno(3.2)
$$
Let $\Omega^{1/2}(x,x')$ be the generalization of an ADF, with the point
$\Lambda$ replaced by an arbitrary point $x' \in M$. (Note that
$\Omega(x,x')$ depends on $n$.) Then
$$
G(x,x') = \Omega^{-1/2}(x,x') + \frac{\mu(x')}{2} + O(\Omega^{1/2}).
\eqno(3.3)
$$
But
$$
\Omega(x,x') = \Omega(x',x)[1 + O(\Omega)] \eqno(3.4)
$$
and, by virtue of the self-adjointness of $L_g$,
$$
G(x,x') = G(x',x), \eqno(3.5)
$$
so that $\mu(x')$ in (3.3) is in fact constant and thus identical with the
ADM mass of the time-symmetric contribution to the initial-data set.
Next, calling
$$
G(x,x') - \Omega^{-1/2}(x,x') = \psi(x,x'), \eqno(3.6)
$$
it follows from the results in [6], that $\lim_{n \ra \infty} \mu_n = \infty$
and there are positive, $n$-independent constants $C$, $C'$ such that
$$
C \mu_n \leq \psi \leq C' \mu_n. \eqno(3.7)
$$
(This is proved in [6] for the Green function $G(x)$, i.e. for $x'$
at $\Lambda$ where $g$, $\partial g$ are kept fixed. It is not
difficult to see that the statement remains true in the present
situation.)

Furthermore, in an arbitrary chart centered at $x$,
$$
D|x-x'|^2 \leq \Omega(x,x') \leq D'|x-x'|^2 \eqno(3.8)
$$
where $|x|^2 = \delta_{ab} x^a x^b$ and $D > 0$, $D' > 0$, both independent
of $n$. Thus
$$
G(x,x') \leq \mbox{const } \left[ \frac{1}{|x-x'|} + \mu_n \mbox{ const}
\right] \eqno(3.9)
$$
and
$$
\mbox{const }\left[ \frac{1}{|x'|} + \mbox{const }\mu_n\right]
\leq G(x'). \eqno(3.10)
$$
Similarly,
$$
\partial G(x,x') \leq \mbox{const }\left[ \frac{1}{|x-x'|^2} + \mu_n
\mbox{ const}\right] . \eqno(3.11)
$$
In order to estimate the integral in Equ. (3.2) for large $n$, it suffices
to take the `worst' part, i.e. the contribution where the integration
is carried out over a neighbourhood $N_\Lambda$ of $\Lambda$ and where
$x$ lies in that neighbourhood (so that both the singularity of
$G(x,x')$ at $x = x'$ and the singularity of $(G^{-7} p_{ab} p^{ab})
(x')$ at $x' = \Lambda$ contribute). Inserting (3.9,10,11) together with
(2.16) into (3.2) and using that the volume element $dV_g(x')$ has a
uniform bound in terms of the Euclidean volume element, we are left with
an integral of the form
$$
I_n(x) = \int_{r' \leq R} \left( \frac{1}{|x - x'|} + \mu_n \mbox{ const}
\right) \left( \frac{1}{r'} + \mu_n \mbox{ const}\right)^{-7}
r'{}^{-9+\ve} r'{}^2 dr' \sin \theta' d\theta' d\vp', \eqno(3.12)
$$
$$
 r < R, \qquad 0 < \ve < 1 ,
$$
where $r'$, $\theta'$, $\vp'$ are to be thought of as polar coordinates
coming from a chart $x'$ centered at $\Lambda$.
After elementary integrations we find that
$$
I_n = O(\mu^{-\ve}), \qquad 0 < \ve < 1. \eqno(3.13)
$$
Similarly, to estimate $\partial h$, it suffices, using (3.9,10,11) and
(2.16), to consider the quantity
$$
J_n(x) = \int_{r' \leq R} \left[ \frac{1}{|x-x'|^2} + \mu_n \mbox{ const}
\right] \frac{r'{}^\ve}{(1 + \mu_n \mbox{ const }r')^7} dr'
\sin \theta' d\theta' d\vp', \qquad 0 < r < R. \eqno(3.14)
$$
We easily see that
$$
J_n(x) = O(r^{\ve - 1}), \qquad 0 < \ve < 1 \eqno(3.15)
$$
Recalling that $\Phi(x) = G(x) + h(x)$ and using (3.7, 3.13, 3.15), it
follows that
$$
\Phi = \Omega^{-1/2} + \frac{\mu_n}{2} + \mu_n O(\Omega^{1/2}) +
O(\mu_n^{-\ve}) \eqno(3.16),
$$
$$
\partial \Phi = \partial (\Omega^{-1/2}) + \mu_n O(1) +
O(\Omega^{(\ve-1)/2}), \eqno(3.17)
$$
where $\Omega^{1/2}$ is an ADF corresponding to the point $\Lambda$.
Now consider level surfaces of $\Omega$.
When $\Omega$ is sufficiently small but positive, these are embedded
two-spheres. Let $n'{}^a$ be the unit normal with respect to the
physical metric $g'_{ab} = \Phi^4 g_{ab}$, pointing towards $\Lambda$.
Then the surface is a TS, provided the quantity $H$ defined by
$$
H =  D'_a  n'{}^a + p'_{ab} n'{}^a n'{}^b \eqno(3.18)
$$
is negative on the surface. We rewrite (3.28) in terms of unbarred
quantities. Using
$$
n'{}^a = - \Phi^{-2} \frac{\Omega^a}{(\Omega_c \Omega^c)^{1/2}},
\eqno(3.19)
$$
we obtain for the first term in (3.18) the expression
$$
- D'_a n'{}^a = \frac{\Omega^{-1/2} \Phi^{-3}}
{(\Omega_c \Omega^c)^{1/2}} \left[ \Omega^{1/2} \Phi
\left( g^{ab} - \frac{\Omega^a \Omega^b}{\Omega_d \Omega^d}\right)
\Omega_{ab} + 4 \Omega^{1/2} \Phi_a \Omega^a \right]. \eqno(3.20)
$$
{}From the definition (2.9) of $\Omega$ and the properties of a CS,
we easily find
$$
\left( g^{ab} - \frac{\Omega^a \Omega^b}{\Omega_d \Omega^d}\right)
\Omega_{ab} = 4 + O(\Omega) \mbox{ uniformly in $n$.} \eqno(3.21)
$$
Using (3.16,17), there results
$$
 D'_a  n'{}^a = \frac{4 \Omega^{-1/2} \Phi^{-3}}{(\Omega_c
\Omega^c)^{1/2}} \left[ 1 - \frac{\mu_n}{2} \Omega^{1/2} +
\mu_n O(\Omega) + O(\Omega^{1/2}\mu_n^{-\ve}) +
O(\Omega^{(1+\ve)/2})\right] . \eqno(3.22)
$$
We now have to estimate the $p'_{ab}$-term in Equ. (3.18). Using
$p'_{ab} = \Phi^{-2} p_{ab}$ and (3.19) we obtain
$$
p'_{ab} n'{}^a n'{}^b = \frac{4 \Omega^{-1/2} \Phi^{-3}}
{(\Omega_c \Omega^c)^{1/2}}
\left[ \frac{1}{4} \frac{\Phi^{-3} \Omega^{1/2}}{(\Omega_d \Omega^d)^{1/2}}
p_{ab} \Omega^a \Omega^b\right]. \eqno(3.23)
$$
Due to (3.16) we have that $\Phi \geq \Omega^{-1/2}$ for large $n$. Using
this fact and (2.16) we see that the bracket in (3.23) is
$O(\Omega^{(1+\ve)/4})$. Combining this estimate with (3.22) it follows
that $H$ is negative provided the quantity
$$
\eta = 1 - \frac{\mu_n}{2} \Omega^{1/2} + A \mu_n \Omega +
B \Omega^{(1+\ve)/4}, \qquad 0 < \ve < 1 \eqno(3.24)
$$
is negative,
where $A$, $B$ are suitable positive constants independent of $n$.
It can be shown that there exist positive constants (independent of
$n$) $\alpha$ and $\beta$ such that $\eta$ is negative for
$$
\frac{2}{\mu_n} + \frac{\alpha}{\mu_n^{3/2}} \leq \Omega^{1/2} \leq
\frac{1}{2A} - \frac{\beta}{\mu_n} \eqno(3.25)
$$
when $n$ is large enough. This completes the proof of our theorem.
\appendix{\section*{Appendix}}

\newcounter{zahler}
\renewcommand{\thesection}{\Alph{zahler}}
\renewcommand{\theequation}{\Alph{zahler}.\arabic{equation}}
\setcounter{zahler}{1}

We do not attempt to formulate a theorem under the most general
conditions. The equation we want to solve is of the following form
\beq
L_g h = (1 + G^{-1}h)^{-7} \rho \qquad \mbox{on }
M' = M \setminus \Lambda
\eeq
where $G^{-1}$ is a positive $C^0$-function on $M$. The source $\rho(x)$
is also positive on $M'$ and smooth. In coordinates $x^a$ centered at
$\Lambda$, it behaves like
\beq
\rho(x) = O(r^{-2 + \ve}), \qquad 0 < \ve < 1,
\eeq
\beq
\partial \rho(x) = O(r^{-3 + \ve}).
\eeq
The metric $g$ on $M$ is such that $\lambda_1(g) > 0$. Hence the operator
$L_g$ in (A.1) has an everywhere positive Green function $G(x,x')$.
Instead of (A.1), we first try to solve equation
\beq
h = \F [h],
\eeq
where $\F[h]$ is defined by
\beq
(\F[h])(x) = \int_M G(x,x')(1 + G^{-1}h)^{-7}(x') \rho(x') dV_g(x').
\eeq
We are thus seeking a fixed point of the mapping $\F$. When $x'$ is in a
local neighbourhood of $x$, $G(x,x')$ satisfies
\beq
G(x,x') = \frac{1}{|x-x'|} + O(1),
\eeq
\beq
\partial G(x,x') = \partial \left( \frac{1}{|x-x'|}\right) +
O\left( \frac{1}{|x-x'|}\right)
\eeq
with $|x-x'|^2 = g_{ab}(x)(x-x')^a(x-x')^b$. (A.6) and the fact that $G > 0$
imply that $\F$ maps  continuous, non-negative functions into themselves.
Let $\psi$ be the continuous function on $M$ defined by
\beq
\psi = \F[0]
\eeq
and $U$ the space
\beq
U = \{ h \in C^0(M)|0 \leq h(x) \leq \psi(x)\}
\eeq
which is clearly a closed, convex subset of the Banach space $C^0(M)$.
Since $\F$ is pointwise decreasing, it maps $U$ into itself. Let $h$ be
in the image of $U$ under $\F$. Then, by (A.7), it is easily seen to be
in $C^1(M \setminus \Lambda)$ with a singularity at $\Lambda$ of the
form
\beq
\partial h = O(r^{\ve -1}).
\eeq
{}From this one easily deduces that $h$ is bounded in $C^{0,1-\ve}(M)$.
Thus $\F(U)$ is pre-compact in $C^0(M)$. Furthermore, for
$h_1, h_2 \in U$, there holds
\beq
| \F[h_1] - \F[h_2]| \leq C|h_1 - h_2|,
\eeq
where $C$ is independent of $h_1$, $h_2$. Thus $\F$, defined on $U$, is
continuous. By the Schauder fixed point theorem (see e.g.
Gilbarg-Trudinger [25], p. 221) it now follows that there exists
$h \in U$, such that
\beq
h = \F [h].
\eeq
Again, this implies that $h \in C^{0,1-\ve}(M)$. Thus the function
$(1 + G^{-1}h)^{-7}$ in (A.5) is Lipschitz continuous. Furthermore,
by virtue of (A.2,3), $\rho$ satisfies
\beq
|\rho(x) - \rho(y)| \leq C \frac{|x-y|}{[\min (|x|,|y|)]^{3-\ve}}
\eeq
near $\Lambda$. Using these facts in (A.5) together with standard potential
theoretic estimates analogous to [25, p. 54] we infer that
$h \in C^2(M \setminus \Lambda)$ and satisfies (A.1). Furthermore
$\Phi = G + h$ obeys the boundary conditions (2.11,12).

Finally, for completeness sake, we show, along the lines of Cantor [26],
that the above solution is unique. Suppose, assuming the contrary, that
there are two solutions, $\Phi_1$ and $\Phi_2$, solving Equ. (2.20).
Then, defining
\beq
\bar g_{ab} = G^4 g_{ab}, \qquad \bar \Phi_i = G^{-1} \Phi_i, \qquad
i = 1,2,
\eeq
\beq
L_{\bar g} \bar \Phi = - \Delta_{\bar g} \bar \Phi = G^{-5} \sigma(x,
\bar \Phi),  \qquad \mbox{on } \wt M
\eeq
where $\sigma$ is a nowhere increasing function of $\bar \Phi$. Thus, with
$\psi$ defined by $\psi = \bar \Phi_1 - \bar \Phi_2$, we have
\beq
- \bar \Delta \psi = \sigma(x,\bar \Phi_1) - \sigma(x,\bar \Phi_2).
\eeq
Suppose $\psi$ is anywhere non-zero, say positive. Then there is an open
set ${\cal O} = \{ x \in \wt M|\psi > 0\}$ which is not the zero-set with
$\psi = 0$ on $\partial {\cal O}$.  Thus
\beq
\bar \Delta \psi \geq 0
\eeq
which, by virtue of the maximum principle, is a contradiction.
Exchanging the role of $\bar \Phi_1$ and $\bar \Phi_2$ we find that, in
fact, $\bar \Phi_1 \equiv \bar \Phi_2$. This ends the uniqueness proof.

\end{document}